# Chemical Migration and Dipole Formation at van der Waals Interfaces between Magnetic Transition Metal Chalcogenides and Topological Insulators

Brenton A. Noesges[1]*, Tiancong Zhu[1], Jacob J. Repicky[1], Sisheng Yu[1], Fengyuan Yang[1], Jay A. Gupta[1], Roland K. Kawakami[1], Leonard J. Brillson[1,2]

[1]*Department of Physics, The Ohio State University, Columbus OH 43210, United States*
[2]*Department of Electrical and Computer Engineering, The Ohio State University, Columbus, OH 43210, United States*

* noesges.1@osu.edu

Metal and magnetic overlayers alter the surface of the topological insulator (TI) bismuth selenide ($Bi_2Se_3$) through proximity effects but also by changing the composition and chemical structure of the $Bi_2Se_3$ sub-surface. The interface between $Bi_2Se_3$ and Mn metal or manganese selenide was explored using x-ray photoelectron spectroscopy (XPS) revealing chemical and electronic changes at the interface. Depositing Mn metal on $Bi_2Se_3$ without an external source of Se shows unexpected bonding within the Mn layer due to Mn-Se bonding as Se diffuses out of the $Bi_2Se_3$ layer into the growing Mn film. The Se out-diffusion is further evidenced by changes in Bi core levels within the $Bi_2Se_3$ layers indicating primarily Bi-Bi bonding over Bi-Se bonding. No out-diffusion of Se occurred when excess Se is supplied with Mn, indicating the importance of supplying enough chalcogen atoms with deposited metals. However, $Bi_2Se_3$ core level photoelectrons exhibited a rigid chemical shift toward higher binding energy after depositing a monolayer of $MnSe_{2-x}$, indicating a dipole within the overlayer. Stoichiometry calculations indicated that the monolayer forms MnSe preferentially over the transition metal dichalcogenide (TMD) phase $MnSe_2$, providing a consistent picture of the dipole formation in which a plane of Se anions sits above Mn cations. This study shows that chemical diffusion and dipole formation are important for Mn-$Bi_2Se_3$ and $MnSe_{2-x}$-$Bi_2Se_3$ and should be considered carefully for TMD/TI interfaces more generally.

## I. INTRODUCTION

Proximity effects between materials can induce new properties at their interface that would not otherwise exist separately. Magnetic overlayers brought into contact with topological insulators (TI) such as $Bi_2Se_3$ can break time-reversal symmetry and affect spin properties leading to quantum phenomena, i.e. the quantum anomalous Hall effect (QAHE) [1,2] and skyrmions [3]. Three-dimensional (3D) ferromagnets such as α-MnSe (111) or even two-dimensional (2D) materials such as 1T-$MnSe_2$ [4] are promising candidates to induce these surface magnetic and spin-related effects in $Bi_2Se_3$ [5]. A crucial component to realizing these advanced interface effects is an abrupt and clean interface between $Bi_2Se_3$ and the ferromagnetic overlayer, particularly for $MnSe_2$ which is a van der Waals (vdW) material with only weak bonding between layers. Molecular beam epitaxy (MBE) growth and subsequent transfers to measurement systems carried out completely in ultra-high vacuum (UHV) provided a method to create and analyze high-quality interfaces without interlayer contamination or surface oxidation. Interlayer contaminants can alter the interface electronic and magnetic properties, and the electronic properties of both constituents will begin to degrade if exposed to oxygen. While UHV helps remove external contaminants, interdiffusion and other chemical interaction at the interface between $Bi_2Se_3$ and magnetic overlayer can still occur. Many metals including Mn have been used to dope or alloy with $Bi_2Se_3$ [6] so that the chemical interaction between the two layers could in principle dominate the electronic and magnetic properties at their interface [7,8]. Furthermore, Mn and $Bi_2Se_3$ can form a magnetic ternary alloy $Bi_2MnSe_4$ which is also a promising system to investigate novel quantum and topological states [9,10,11].

Surface-sensitive techniques such as XPS can provide insight into the impact of Mn-containing overlayers on the electronic structure of $Bi_2Se_3$ and how Mn incorporates with $Bi_2Se_3$ at the $MnSe_2$ interface. We used XPS connected via ultra-high vacuum (UHV) suitcase to a MBE system to investigate chemical interaction at the interface between a selenide TMD and $Bi_2Se_3$. We compared the chemical effects on pristine $Bi_2Se_3$ of Mn-containing overlayers in two different scenarios: (1) extreme Se-deficient case, i.e., Mn deposited alone onto $Bi_2Se_3$ and (2) Mn deposited with excess Se. In the case of pure Mn metal on $Bi_2Se_3$ we saw that Mn reacts with the $Bi_2Se_3$ surface where Mn bonds with Se extracted by diffusion from the $Bi_2Se_3$ substrate. In contrast, when excess Se was provided, there was little chemical interaction observed at the interface between $MnSe_{2-x}$ and $Bi_2Se_3$. Instead, $MnSe_{2-x}$ on $Bi_2Se_3$ rigidly shifted subsurface $Bi_2Se_3$ core levels toward higher binding energy indicating the formation of a negative surface/interface dipole. The presence of this dipole is likely due to growth of primarily α-MnSe instead of the $1T-MnSe_2$ 2D phase which is supported by topographic scanning tunneling microscopy (STM) images and spectroscopy. Our XPS core level analysis combined with controlled depositions, air-free transfers and surface analysis provides a consistent picture of chemical diffusion and dipole formation at the Mn selenide/$Bi_2Se_3$ junction, revealing interfacial effects that may be relevant to TMD/TI heterostructures in general.

## II. METHODS

*Molecular beam epitaxy (MBE)* – Prior to $Bi_2Se_3$ sample growth, $Al_2O_3(0001)$ substrates were annealed in the air at 1000°C for 3 hours. The substrates were then transferred to a home built molecular beam epitaxy (MBE) chamber and heated to the growth temperature of 250°C. Bi and Se were evaporated from standard Knudsen cells. A Se/Bi flux ratio (by thickness) of 20:1 was used during the growth with Bi cell at 575°C and Se cell at 180°C. $Bi_2Se_3$ was deposited at a growth rate of 1 nm/min. After the growth, the samples were cooled down to room temperature in vacuum. A 20 nm Se capping layer was deposited on top of $Bi_2Se_3$ to prevent oxidation before taking the sample out of the chamber.

Both the Mn-$Bi_2Se_3$ and the $MnSe_{2-x}$-$Bi_2Se_3$ were grown in a separate MBE chamber (Veeco GEN930). The Se-capped $Bi_2Se_3$ samples were mounted onto 18 mm flag style paddles and loaded into the MBE chamber using special adapters to the 75 mm uniblock sample plates. The samples were first gently annealed at 170°C to remove the Se capping. Reflection high energy electron diffraction (RHEED) was used to monitor the sample surface and ensure that the cap removal is complete. During deposition, both Mn and Se were evaporated from a Knudsen cell. For the Mn-$Bi_2Se_3$ sample, Mn was deposited at room temperature with a beam flux of $5 \times 10^{-9}$ torr. For the $MnSe_{2-x}$-$Bi_2Se_3$ sample, Mn and Se were co-deposited at 250°C with a flux ratio of ~1:50 (by beam equivalent pressure).

*In vacuo Transfer and XPS measurement* – To prevent oxidation, all samples were transferred between the $MnSe_{2-x}$ growth chamber and XPS chamber using an ultrahigh vacuum (UHV) suitcase tool which carries the 18 mm sample paddles. XPS measurements provided chemical composition and bonding information for the de-capped $Bi_2Se_3$ surface and two overlayer samples, 1.5 nm of Mn metal and 1 monolayer (ML) of $MnSe_{2-x}$.

XPS was performed using a PHI VersaProbe 5000™ system equipped with a Scanning XPS Microprobe X-ray source ($h\nu_{Al\ K\alpha}$ = 1486.6 eV; FWHM ≤ 0.5 eV), and hemispherical energy analyzer with a pass energy of 23.5 eV and 0.05 eV step. To minimize the effects of charging, the XPS system is equipped with a two-stage sample surface neutralization system consisting of a 10 eV electron flood gun accompanied by a 10 eV $Ar^+$ ion beam. Photoelectrons were collected at a takeoff angle of 45°. Chemical composition was determined using the PHI MultiPak analysis suite to determine relative atomic concentration using relative sensitive factors [12] for Bi 5d, Bi 4f, Se 3d, and Mn 2p core levels of 26.089, 180.178, 14.492, and 53.366 respectively. Additional curve fitting was performed using asymmetric Gaussian or Voigt line shapes for metallic and non-metallic states respectively.

*STM measurement* – STM measurements were performed at 5 K using a CreaTec LT-STM. Samples were transferred from the MBE growth chamber to the STM system through a UHV suitcase to preserve the as-grown quality of the films. No additional preparation of the surface was performed prior to measurement. Images were obtained in constant current mode with a setpoint of 50 – 500 pA and sample bias in the range of 0.5 – 1 V. Tunneling spectra were obtained by adding a 20-50 mV modulation voltage at 1063 Hz to the DC bias and measured using a lock-in amplifier. Image processing and analysis were performed using WxSM [13].

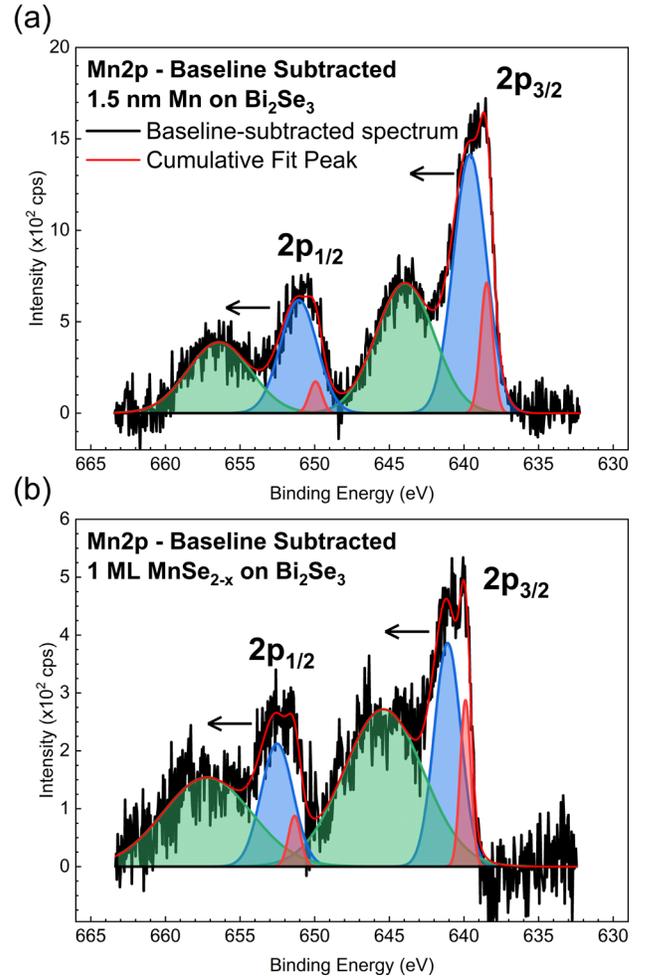

|  | $Bi_2Se_3$ | $Mn/Bi_2Se_3$ | $MnSe_{2-x}/Bi_2Se_3$ |
|---|---|---|---|
| % Bi | 50.9±0.7 | 45.0±1.0 | 40.9±0.6 |
| % Se | 49.1±0.8 | 31.0±1.0 | 49.2±0.7 |
| % Mn | -- | 24.5±0.6 | 9.9±0.5 |
| Bi:Se Ratio | 1.03±0.02 | 1.47±0.06 | Assuming 1.03 |
| Mn:Excess Se Ratio | -- | -- | 1.0±0.1 |

TABLE I – Stoichiometry found by determining relative peak areas below Se 3d, Mn 2p and an average of the Bi 4f and Bi 5d core levels. Peak areas were normalized with relative sensitivity factors appropriate for each core level. Mn:Se ratio was estimated for the case of $MnSe_{2-x}$ on $Bi_2Se_3$ by assuming the Bi:Se ratio as found in the bare $Bi_2Se_3$ surface.

FIG. 1. Mn 2p core level spectrum of deposited Mn metal indicating unexpected Mn-Se bonding as shown by satellite features in green shifted to higher binding energy as indicated by the arrows. (b) $MnSe_{2-x}$ Mn 2p core level showing similar satellite features indicating Mn-Se bonding.

### III. RESULTS

***Bare $Bi_2Se_3$ Surface*** – The bare surface of de-capped $Bi_2Se_3$ was first studied to check for surface contamination and establish a baseline for Bi and Se XPS peak centers and stoichiometry. Wide energy range XPS surveys and detailed region scans showed no detectable core level signals other than the expected Bi and Se levels and related Auger electron features. The de-capped surface showed sharp Bi 5d, Bi 4f and Se 3d doublets with binding energies (BE) consistent with previously measured $Bi_2Se_3$ core levels [14,15]. The stoichiometry of the $Bi_2Se_3$ surface (Table I) was obtained by comparing the relative peak areas below the Bi 5d, Bi 4f and Se 3d spectra where the corrected peak areas for Bi 5d and 4f were averaged together. The Bi:Se ratio was found to be 1.03±0.02 indicating a Se-deficient surface (Table I). The XPS spectra also show a pair of small broad features near the 285 eV region where a C 1s line would be (Supplemental Material, SM, Fig. S1(a)). However, neither feature is a C 1s feature but instead they are related to Se Auger electron transitions (SM Fig. S1(b)). Since the O 1s line was below the detection threshold of the survey scan, a detailed region scan was not performed. The lack of surface contamination of the samples transferred via UHV suitcase confirms the quality of the suitcase transfer and its ability to maintain a clean, oxide-free sample surface.

***Mn – $Bi_2Se_3$*** – To study the interaction of Mn metal with the $Bi_2Se_3$ surface, we deposited Mn metal only, e.g., without Se, onto $Bi_2Se_3$. The wide range survey scan (SM Fig. S1(a)) shows virtually identical peaks to the bare $Bi_2Se_3$ surface with the addition of Mn-related peaks, notably the Mn 2p core level. Detailed region scans of the Mn 2p core level (Fig. 1(a)) show additional satellite features shifted to higher BE relative to the metallic Mn peak features. The shift to higher BE indicates Mn-Se bonding as evidenced by the increase in BE and charge state of the Mn ions. Additionally, the Mn 2p line shape is consistent with past measurement and modeling of manganese octahedrally coordinated to form $MnX_6$ clusters [16,17].

There are two likely sources for the Se observed in the Mn metal overlayer: (1) Se pulled from the $Bi_2Se_3$ layers below or (2) residual Se from the growth chamber. The Bi core levels provides insight into the source of this Se in the Mn layer. Both Bi 5d and 4f core levels (Fig. 2(a) & 2(b)) exhibit a 1.74 eV shift toward lower BE after Mn metal was deposited. However, such a large negative BE shift is not seen in the Se 3d core level (Fig. 2(b)) This large negative BE shift in bismuth core levels has been previously identified as $Bi_2$ layer formation within the bismuth selenide [14,18]. A careful deconvolution of the Bi 4f core level spectra (Fig. 3) reveals a small feature at higher BE compared to the dominant Bi $4f_{7/2}$ peak. This smaller feature has BE consistent with the $Bi_2Se_3$ previously observed indicating that Mn is not fully converting the subsurface $Bi_2Se_3$ into bismuth bilayers. Such changes in the bismuth core levels were not observed in earlier XPS studies on Mn-doped $Bi_2Se_3$ [19]. This is evidence that the Se in the Mn layer has diffused out of the $Bi_2Se_3$ leaving behind $Bi_2$ bilayers rather than Se pulled from the surrounding chamber.

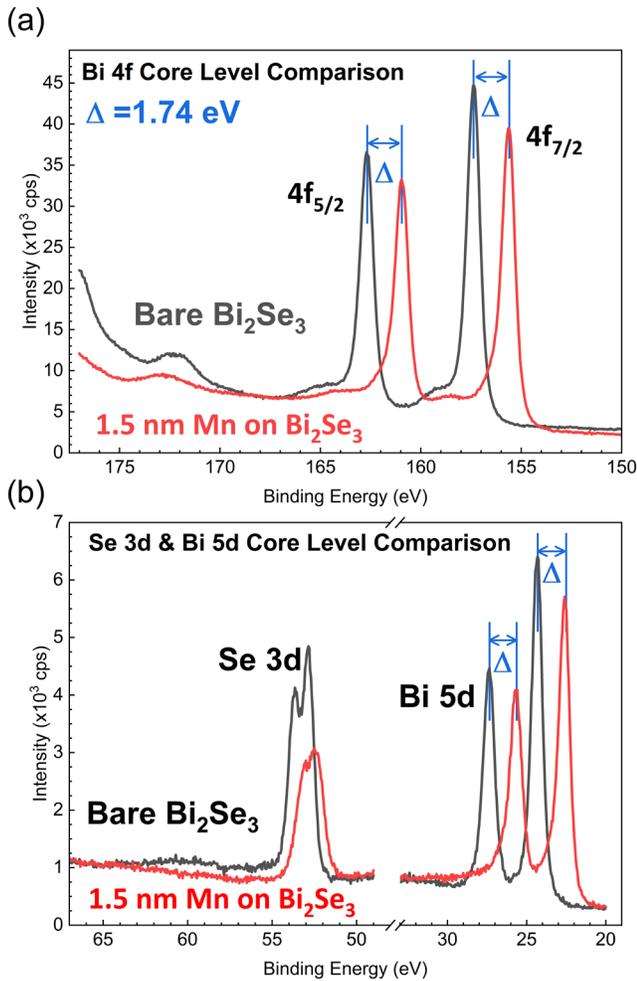
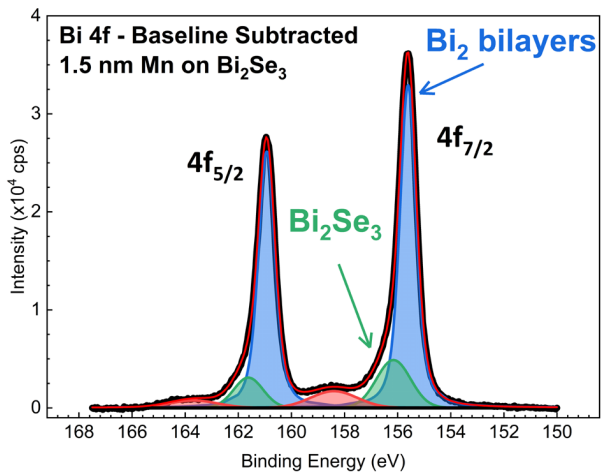

FIG. 2. (a) Bi 4f core level showing the binding energy difference before (black) and after (red) 1.5 nm Mn deposition. Depositing 1.5 nm of Mn causes Bi core levels of $Bi_2Se_3$ to shift toward higher binding energy. (b) Se 3d and Bi 5d core levels. The Se 3d peaks do not exhibit the same 1.74 eV shift as the Bi core levels.

FIG. 3. Deconvolved Bi 4f core level spectra for 1.5 nm of Mn metal on $Bi_2Se_3$. The dominant peak at lower binding energy is related to the formation of $Bi_2$ layers within $Bi_2Se_3$

**$MnSe_{2-x}$ – $Bi_2Se_3$** – There is a very different interface picture on the side of $Bi_2Se_3$ when Mn is deposited with excess Se in a growth targeted for one unit-layer of $MnSe_2$. Fig. 1(b) shows that the Mn 2p core level of $MnSe_{2-x}$ shows a similar peak structure but shifted 1.4 eV toward higher BE compared to that of the deposited Mn metal on $Bi_2Se_3$ in Fig. 1(a). The similar deconvolved peak structure confirms that the Mn metal BE shown above is consistent with Mn-Se bonding. However, all the photoelectron BE measured from $Bi_2Se_3$ after depositing 1 ML $MnSe_{2-x}$ exhibit a rigid 0.8 eV chemical shift toward higher BE compared to bare $Bi_2Se_3$ (Fig. 4 (a) and (b)). This shift toward higher observed BE indicates the presence of a negative interface dipole. Given the increase in BE, such a dipole represents an extra potential barrier at the free surface that escaping core level photoelectrons most overcome shown schematically in Fig. 5.

We used the relative peak areas below deconvolved Mn 2p, Se 3d, and averaged Bi 4f and 5d core level features to calculate the surface stoichiometries shown in Table I. Based on the Bi:Se ratio of 1.03 as observed for the bare $Bi_2Se_3$, the ratio of Mn:Se is 1.0±0.1, indicating a Se deficiency and the chemical phase of the manganese selenide overlayer is more likely MnSe instead of $MnSe_2$. The formation of MnSe versus the van der Waals material MnSe$_2$ helps explain the origin of the observed dipole. A single monolayer of MnSe would form a negatively charged plane of Se ions on positively charged Mn creating the dipole that retards exiting electrons and causes the core level shift to higher BE.

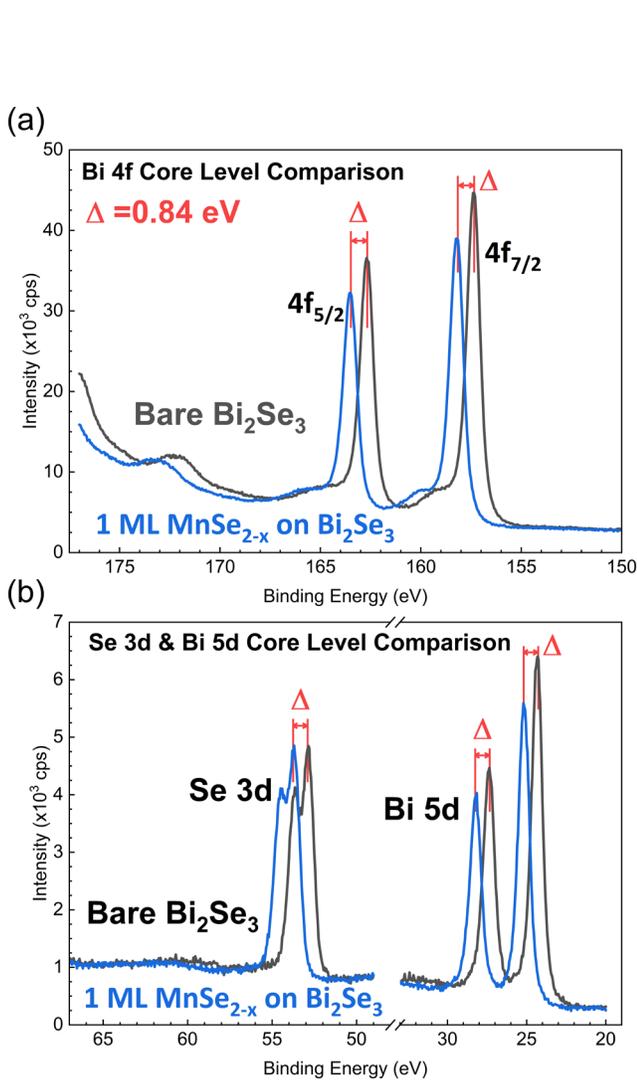

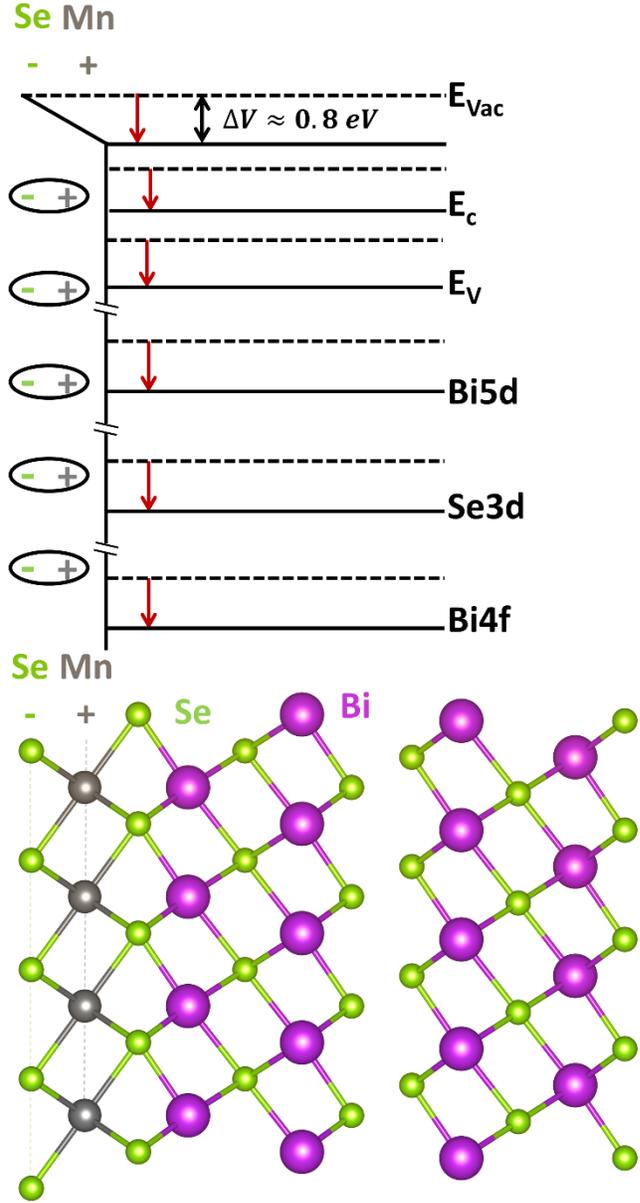

FIG. 4. (a) Bi 4f core level showing the binding energy difference before (black) and after (blue) 1 ML MnSe$_{2-x}$ deposition. (b) Se 3d and Bi 5d core levels. The Se 3d peaks do not exhibit the same 1.74 eV shift as the Bi core levels. Depositing 1 ML of MnSe$_{2-x}$ causes the Bi and Se core levels to rigidly shift 0.84 eV toward higher binding energy.

FIG. 5. Schematic band and lattice diagram showing the location of expected dipole and effect on the Bi$_2$Se$_3$ core levels below the TMD/TI interface. Crystal structure produced using VESTA [20].

Consistent with these XPS results, STM surface characterization suggests that $\alpha$-MnSe(111) is the dominant phase within the MnSe$_{2-x}$ interfacial region. Atomic resolution of a typical $\alpha$-MnSe(111) island is shown in Fig. 6(a). The inset shows a Fourier transform of the area confirming well-ordered structures with triangular symmetry and a lattice constant of 3.97$\pm$0.04Å consistent with $\alpha$-MnSe(111) [21]. The topographic line profiles and dI/dV spectroscopy (Fig. 6(c) and (d)) indicate that $\alpha$-MnSe(111) islands have an apparent height of 2.9±0.2Å and an energy gap of 700 meV. The observed gap is smaller than the 3 eV gap expected for bulk $\alpha$-MnSe [21] but the measured step height and atomic lattice constant is in good agreement with other measurements of α-MnSe(111) [8], consistent with our expected XPS stoichiometry. Fig. 6(b) also shows that a small secondary phase distinct from $\alpha$-MnSe(111) exists on the surface. In comparison to α-MnSe, these small (< 10 nm diameter) islands show apparent heights of 5.8±0.1Å and exhibit a distinct dI/dV spectrum (Fig. 6d) with a much smaller gap. This is suggestive of 1T-$MnSe_2$ formation, but these smaller secondary phase regions are too small and low density to measure with XPS for further characterization.

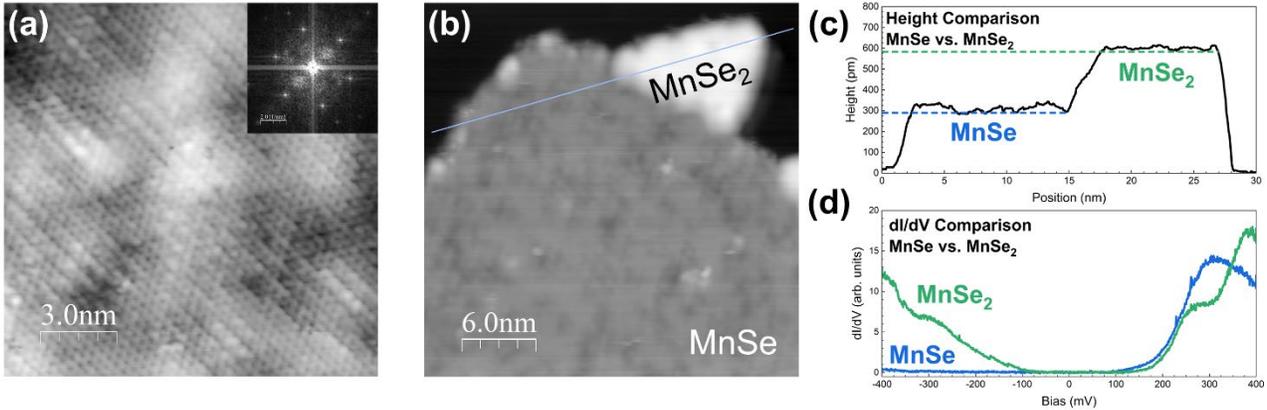

FIG. 6. (a) Atomically resolved STM image of a $\alpha$-MnSe(111) island on $Bi_2Se_3$ (10 mV, 500 pA). Inset shows a Fourier transform image of the area with a lattice constant of 3.97±0.04Å. (b) Topographic image of a 1T-$MnSe_2$ island adjacent to a $\alpha$-MnSe(111) island (1 V, 50 pA). (c) Line profile of the two islands measured along the blue line in (b). The dashed line represents the average height of the step for MnSe (blue) and $MnSe_2$ (green) (d) Comparison of dI/dV spectroscopy collected on $\alpha$-MnSe(111) (blue) and 1T-$MnSe_2$ (green).

## IV. ANALYSIS

*Mn – $Bi_2Se_3$* – Pure Mn metal interacts strongly at the interface with $Bi_2Se_3$ where Se diffuses out of $Bi_2Se_3$ into the Mn overlayer creating clusters of $MnSe_{2-x}$ as indicated by the satellite features in the Mn 2p core level (Fig. 1a). The removal of Se from $Bi_2Se_3$ also leaves behind metallic Bi bilayers below the interface as evidenced by the strong 1.74 eV negative binding energy shifts of the Bi 5d and 4f core levels. This Se out-diffusion has been observed in other $Bi_2Se_3$-metal interfaces with Cr and Fe overlayers [15]. The difference in chemical reactivity between MnSe and $MnSe_2$ may also factor into the MnSe formation. The chemical equations for Mn reaction at the surface of $Bi_2Se_3$ are Mn+ (1/3) $Bi_2Se_3$ → MnSe+ (2/3) Bi, for MnSe and Mn+ (2/3) $Bi_2Se_3$ → $MnSe_2$+ (4/3) Bi, for $MnSe_2$. The heat of reaction $\Delta H_R$ per metal atom for MnSe ($H_F$ = -155.6 kJ/mol) [22] versus $Bi_2Se_3$ ($H_F$ = -140.2 kJ/mol) [23] is -108.8 kJ/mol or -1.12 eV/metal atom compared versus $MnSe_2$ ($H_F$ = -180.5 kJ/mol) with $Bi_2Se_3$ that yields -87 kJ/mol or - 0.90 eV/metal atom, which is less reactive.

*$MnSe_{2-x}$ – $Bi_2Se_3$* – There is less chemical reactivity at the interface when Se is supplied alongside Mn during deposition. There is no evidence of Se outdiffusion or metallic bismuth formation below the $MnSe_{2-x}$/$Bi_2Se_3$ interface as there was in the case of Mn metal. However, the existence of a single monolayer of MnSe$_{2-x}$ produces a 0.74 eV rigid shift toward higher binding energy observed in Bi$_2$Se$_3$ core level photoelectrons. This increase in binding energy is due to an interface dipole created by the dominant manganese mono-selenide phase that formed over the MnSe$_2$ phase. α-MnSe (111) forms a plane of negatively charged Se ions at the free surface sitting above positive Mn ions which creates the observed negative dipole that slows escaping core level photoelectrons. Smaller regions of MnSe$_2$ also exist on the Bi$_2$Se$_3$ surface but these regions are too small and low density to compare results between MnSe and MnSe$_2$ regions. Presumably, such a dipole would not exist in a predominantly MnSe$_2$/Bi$_2$Se$_3$ interface.

## V. CONCLUSIONS

We have used surface science techniques to explore the interface chemistry between Bi$_2$Se$_3$ and Mn-containing overlayers and found surface reactions and diffusion between constituents that depends significantly on the Se conditions during deposition. Mn metal deposited without excess Se induces Se out-diffusion from the Bi$_2$Se$_3$ substrate. This behavior suggests that Mn should be supplied with excess selenium to avoid altering the Bi$_2$Se$_3$ on which the Mn is deposited. When Se is supplied with Mn, no outdiffusion of Se occurs. Rather, Bi$_2$Se$_3$ core level photoelectrons exhibit a rigid shift toward higher binding energy indicating a surface dipole caused by the formation of primarily α-MnSe (111) with smaller regions of MnSe$_2$. The existence of this dipole can be useful to indicate the phase of the grown manganese selenide overlayer when exact stoichiometry analysis is challenging or impossible due to a shared core level. Further, this work points to the importance of supplying excess chalcogen during deposition to avoid chemical bonding with substrate chalcogen atoms. This evidence for chemical diffusion and dipole formation at the Mn selenide/Bi$_2$Se$_3$ junction reveals interfacial effects that may be relevant to TMD/TI heterostructures in general.

## ACKNOWLEDGEMENTS

This research was primarily supported by the Center for Emergent Materials: an NSF MRSEC under award number DMR-1420451 (XPS measurements by BAN and LJB). TZ, JJR, JAR, and RKK acknowledge support from the US Department of Energy (DOE), Office of Science, Basic Energy Sciences under Grant No. DE-SC0016379 (growth of Mn and Mn-selenide films and STM measurements). SY and FY acknowledge support from the US DOE, Office of Science, Basic Energy Sciences, under Grant No. DE-SC0001304 (growth of Bi$_2$Se$_3$ films).

# Supplemental Material

# Chemical Migration and Dipole Formation at van der Waals Interfaces between Magnetic Transition Metal Chalcogenides and Topological Insulators

Brenton A. Noesges[1], Tiancong Zhu[1], Jacob J. Repicky[1], Sisheng Yu[1], Fengyuan Yang[1], Jay A. Gupta[1], Roland K. Kawakami[1], Leonard J. Brillson[1,2]

[1]*Department of Physics, The Ohio State University, Columbus OH 43210, United States*
[2]*Department of Electrical and Computer Engineering, The Ohio State University, Columbus, OH 43210, United States*

## Table of contents

Figure S1 – XPS Survey spectra and detail scans near C 1s region

## Survey and C 1s Region

Wide range XPS survey scan showing only Mn, Bi and Se related features. Features appearing around 285 eV are related to Se Auger electron features rather than signal due to adventitious carbon.

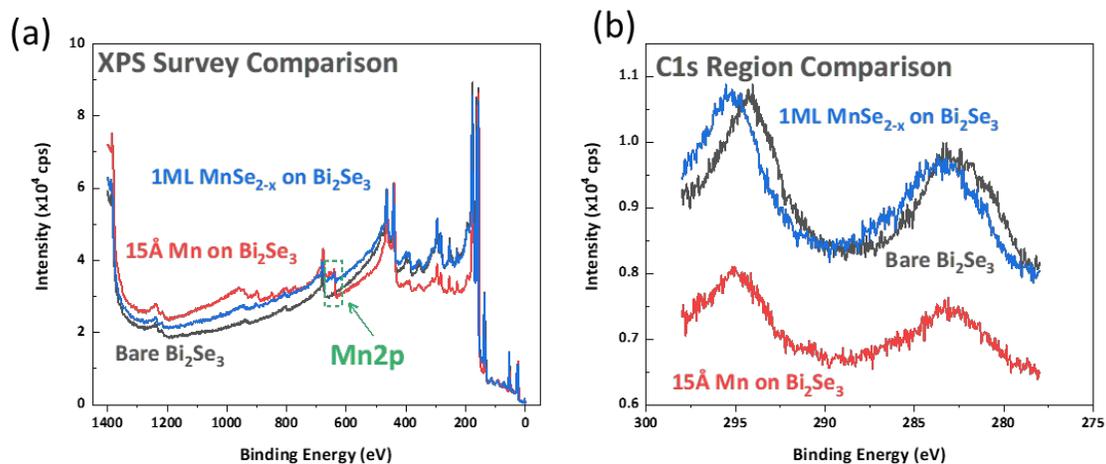

FIG S1. (a) Wide range XPS surveys showing only Mn, Bi and Se related core levels and features. (b) Detailed region scans near C 1s region showing only features indicative of Se Auger electrons and not adventitious carbon.